\documentclass[12pt,preprint]{aastex}

\newcommand{\xmm}{XMM-{\em Newton}}

\newcommand{\msim}{\raisebox{-.4ex}{$\stackrel{>}{\scriptstyle \sim}$}}

\newcounter{ion} \newcommand{\eli}[2]{\setcounter{ion}{#2}#1{~\sc\roman{ion}}}

\def \etal   {\hbox{et~al.\/}}

\def\changed{}

\shorttitle{First high-resolution X-ray spectrum of a single WR star}
\shortauthors{Oskinova et al.}

\begin{document}
\title{High-resolution X-ray spectroscopy reveals the special
nature of Wolf-Rayet star winds}


\author{L.M. Oskinova}
\affil{Institute for Physics and Astronomy, University Potsdam, 
14476 Potsdam, Germany\\
\email{lida@astro.physik.uni-potsdam.de}}
\author{K.G. Gayley}
\affil{Department of Physics and Astronomy, University of Iowa, 
Iowa City, IA 52245, USA}
\author{W.-R. Hamann} 
\affil{Institute for Physics and Astronomy, University Potsdam, 
14476 Potsdam, Germany}
\author{D.P. Huenemoerder}
\affil{Massachusetts Institute of Technology, Kavli Institute 
for Astrophysics and Space Research, 70 Vassar St., Cambridge, 
MA 02139, USA}
\author{R. Ignace}
\affil{Department of Physics and Astronomy, East Tennessee State 
University, Johnson City, TN 37663, USA}
\author{A.M.T. Pollock}
\affil{European Space Agency XMM-Newton Science Operations Centre, 
European Space Astronomy Centre, Apartado 78, 
Villanueva de la Ca\~nada, 28691 Madrid, Spain}



\begin{abstract}    
We present the first high-resolution X-ray spectrum of 
a putatively single Wolf-Rayet star.  400\,ks  observations of
WR\,6 by the \xmm-telescope resulted in a superb quality high-resolution 
X-ray spectrum. 
Spectral analysis reveals that the X-rays originate   far out in the
stellar wind,  more than 30 stellar radii from the  photosphere, and
thus outside the wind acceleration zone where  the line-driving
instability could create shocks. The X-ray emitting plasma reaches
temperatures up to 50\,MK, and is embedded within the un-shocked,
``cool'' stellar wind as revealed by characteristic spectral signatures. 
We detect a fluorescent Fe line  at $\approx 6.4$\,keV. The presence
of fluorescence is consistent with a two-component medium, where
the cool wind  is permeated with the hot X-ray emitting plasma.   
The wind must have a  very porous structure to allow the observed 
amount of X-rays to escape. We find that neither the  line-driving 
instability nor any alternative binary scenario can explain the data.
We suggest a scenario where X-rays are produced when the fast wind rams 
into slow ''sticky clumps'' that resist acceleration. Our new data show 
that the X-rays in single WR-star are generated by some special mechanism 
different from the one operating in the O-star winds.

\end{abstract}



\section{Introduction}

Massive stars reach the Wolf-Rayet (WR) evolutionary phase when
their hydrogen fuel has been consumed or lost and the products of
nuclear fusion appear in their atmospheres, {\changed before ending 
their short lives among the progenitors of core-collapse supernovae 
(SNe) \citep{smartt2009}}. These stars are millions
of times more luminous than the sun and drive strong stellar winds by
radiation  pressure exerted through absorption and scattering in
spectral lines  \citep{nugis2002,graf2005}. 
This ``line-driving'' mechanism is known to be  unstable \citep{lw1980}
and is thought to create hydrodynamic shocks that heat the plasma to a
few million degrees with the emission of thermal X-rays \citep{feld1997}. 
For O-type stars, which have weaker winds than WR stars, this model is
largely consistent with the observations \citep{osk2006,gn2009}. 
The winds of WR stars are also expected to suffer from  the line-driving
instability (LDI) \citep{go1995} and thus may emit  X-rays similar
to the O-type stars. This conjecture was not  verified by observations 
so far. 

In this {\em Letter} we report the first high-resolution 
X-ray spectrum of a putatively single WN-type star and 
its analysis.  

\section{The WN star WR\,6}
\label{sect:thestar}

The object of our study, WR\,6 (EZ\,CMa, HD\,50896),  has been
successfully modeled as a hydrogen-free WN star of subtype WN5 by
fitting its UV, visible, and infrared spectra \citep{wrh2006}. Its
stellar temperature is 90\,kK,  and its luminosity is
$10^{5.6} L_\odot$ when adopting a distance of
1.82\,kpc as implied from its membership to the association
Collinder\,121. The mass-loss rate is
$\dot{M}\approx 2\times 10^{-5}\,{M_\odot}$\,yr$^{-1}$, and the wind reaches a
terminal velocity of $v_\infty \approx 1,700$\,km\,s$^{-1}$.  Hence the
wind carries a mechanical power $\dot{M} v^2_\infty / 2$ equivalent to
$10^4L_\odot$, which is about 3\% of the stellar luminosity.

The WR winds are strongly inhomogeneous \citep{lep1999}. Narrow spectral
features drifting over the broad emission-line profiles are interpreted
as showing radially accelerated blobs of matter \citep{mof1988}.
Quasi-periodic variability could be related to rotation  \citep{dess2002}
and caused by spiral-like density patterns in the wind rooted in the 
photosphere, similar to the corotating interaction regions
observed in the solar wind \citep{mul1984,morel1997}. WR\,6 shows
considerable photometric, spectral, and polarimetric variability on the 
time scale of 3.766\,d \citep{duij1996,stlou2009}.

Its thermal radio spectrum \citep{doug2000} and complex variability 
pattern \citep{duij1996} support the conclusion that WR\,6 is not a binary
but a single star.  Previous X-ray observations of WR\,6 at lower
spectral resolution  confirmed that its X-ray luminosity and temperature
are typical of single WN stars \citep{pol1987,ign2003,osk2005,sk2010}.

\section{Observations and Spectral Modeling}
\label{sect:obs}

The X-ray data on WR\,6  were taken with the X-Ray  Multi-Mirror 
Satellite \xmm.  Its telescopes illuminate three different  instruments 
which always operate simultaneously: RGS is a Reflection Grating 
Spectrometer, achieving a spectral resolution of 0.07\,\AA; RGS is not
sensitive for wavelengths shorter than 5\,\AA. The other 
focal instruments MOS and PN cover the shorter wavelengths; their spectral 
resolution is modest ($E/\Delta E\approx $20 -- 50).

The data were obtained at four epochs in 2010 (Oct.\,11 and 13, Nov.\,4
and 6). The total exposure time of 400 kilosecond was  split into
four individual parts. Our data reduction involved standard procedures of 
the \xmm\ Science Analysis System v.10.0. 

\begin{figure*}
\centering
\includegraphics[width=1.0\columnwidth]{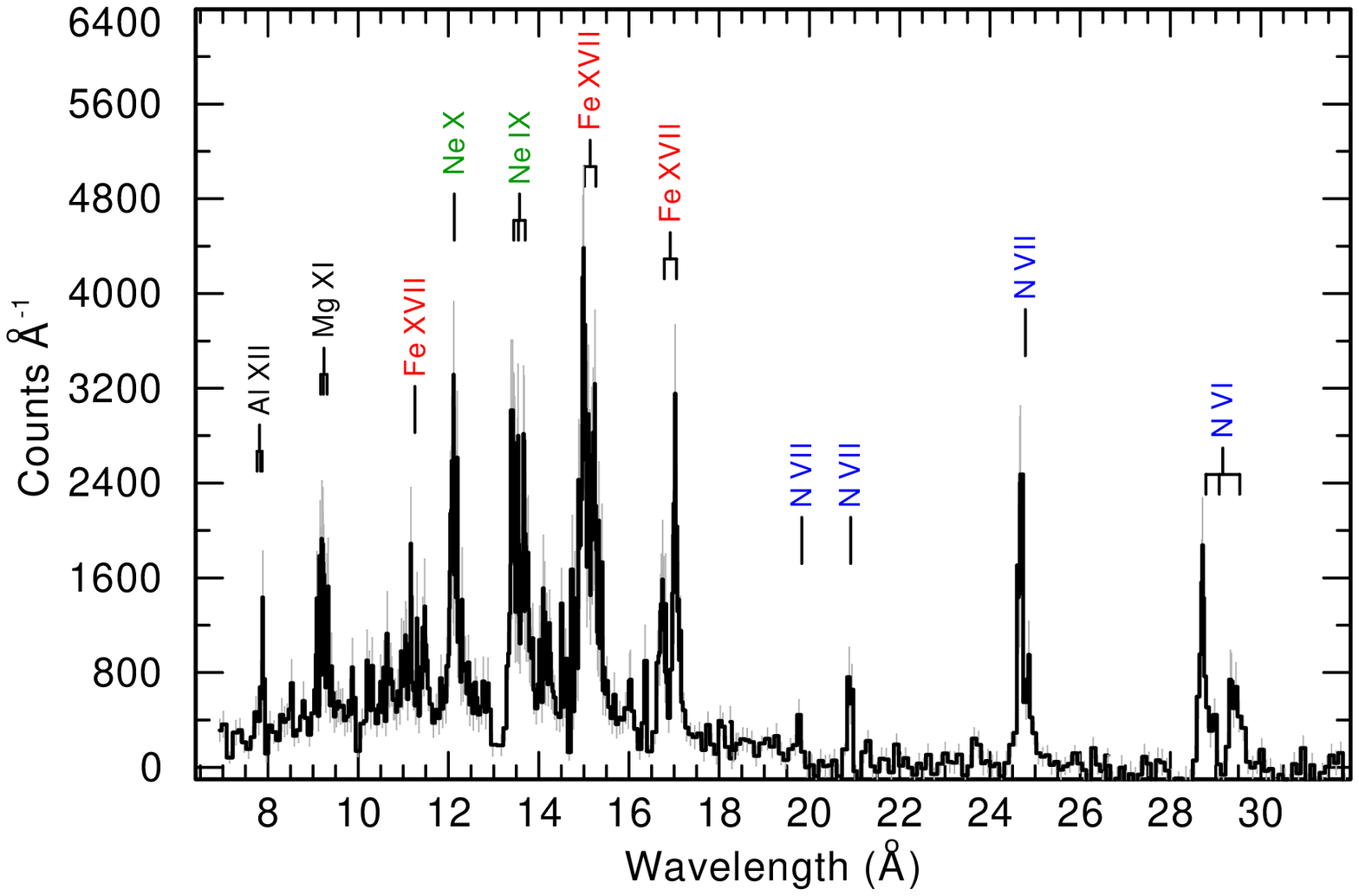}
\caption{The RGS spectrum of WR\,6 with the strongest emission 
lines identified. The error bars correspond 3$\sigma$. The RGS\,1
spectrum is shown from 18.2\,\AA\ to  26.7\,\AA, otherwise the RGS\,2
spectrum is plotted}
\label{fig:rgs4} 
\end{figure*}

The RGS spectrum is shown in Fig.\,\ref{fig:rgs4}. The X-ray luminosity of 
WR\,6  in the 0.3 -- 12.0\,keV band is  $L_{\rm X}\approx 8 \times
10^{32}$\,erg\,s$^{-1}$, or $0.1L_\odot$.    This constitutes $10^{-5}$
of the wind's mechanical power. The RGS  spectrum is dominated by strong
and broad emission lines of metals  in accordance with the WN-wind
abundances characteristic of CNO processed  material. 

To model the observed X-ray spectrum we generated emissivities for a 
hydrogen-free plasma with the APEC thermal plasma model \citep{Smith:01}.  
For most elements the abundances were fixed to the values derived from  
the UV and optical spectra. 

In this stage we approximate the line profiles by Gaussians, with a
common Doppler shift against the laboratory wavelength of
$-650\,\mathrm{km\,s^{-1}}$ and a width (FWHM) of
$3000\,\mathrm{km\,s^{-1}}$ as the best values found by the automatic
procedure from fitting the RGS data. The optimum fits to the observed 
spectra were obtained with the ISIS software \citep{Noble:Nowak:2008}.  

\begin{figure*}
\centering
\includegraphics[width=.7\columnwidth]{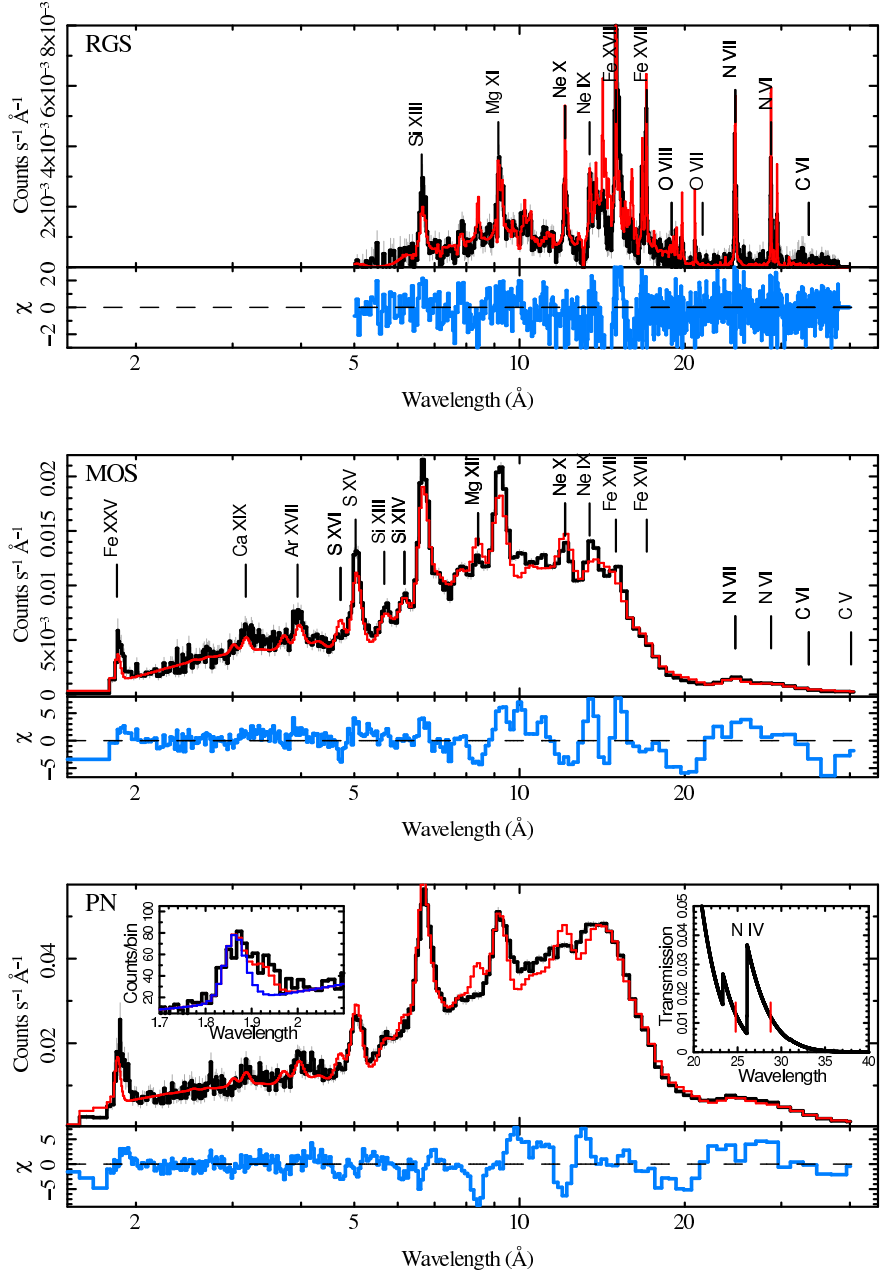}
\caption{The X-ray spectrum of WR\,6, obtained with different
detectors: RGS (top panel), MOS (middle), and PN (bottom). 
Black curves display the observations; the model is shown in red.  
The lower part of each panel displays the residuals. The PN panel 
includes an inset to the right showing the transmission of the cool wind plus 
interstellar absorption for $\lambda > 20\,\mathrm\AA$. The \eli{N}{4} K-shell 
absorption edge is located between the \eli{N}{6} and \eli{N}{7} lines (red 
vertical bars). An insert to the left in the  
PN panel shows a broad spectral feature that is not reproduced by
the thermal plasma model (blue dotted line). An additional
line at 1.92\,\AA, which we identity as fluorescent Fe\,K$\alpha$, 
improves the fit (red histogram).} \label{fig:specmodels} 
\end{figure*}

The observed spectra, the best-fitting model, and the residuals  are
shown in  Fig.\,\ref{fig:specmodels}. In order to achieve an acceptable 
fit, it was necessary to compose the emission from plasmas of three 
different temperatures (1.6, 7.0 and 45\,MK) with respective emission 
measures of 253, 86, and 22 $\times 10^{54}\,\mathrm{cm^{-3}}$. The 
hottest component is required to reproduce the 2--5$\,$\AA\ continuum 
and the emission lines from \eli{Fe}{25}, \eli{Ca}{19}, \eli{Ar}{17}, 
and \eli{S}{15}. Nevertheless, there are still residuals at some lines, 
such as \eli{Mg}{12} and \eli{Ne}{10}, indicating that a three-temperature 
model is not fully sufficient, or that  some abundances are not well
determined. O and C lines are absent or very weak, as expected for a
plasma with the chemical composition of a WN-type wind.

To account for the absorption within the wind we applied the  
{\em vphabs} model \citep{Arnaud:1996}, modified  for the absence of hydrogen. 
Strong absorption, significantly exceeding that of the interstellar medium 
towards the star, is evident. The ratio of fluxes in the N\,{\sc vi} and 
N\,{\sc vii} Ly$\alpha$ lines
is reproduced when absorption in the cool wind is included, especially the 
K-shell ionization edge of N\,{\sc iv} that is located between the two 
aforementioned lines (see Fig\,\ref{fig:specmodels}).

The low-resolution EPIC spectra provide data up to 12\,keV. 
Significant residuals are encountered in the region around the \eli{Fe}{25} 
line at about 1.9\,\AA\ 
(see Fig.\,\ref{fig:specmodels}). The observation shows a 
definitely broader feature than reproduced by the plasma model.  Obviously, 
this complex includes a further blend. The fit of an additional gaussian 
profile yields a flux of $4.5\times10^{-7}\,\mathrm{photons\,cm^{-2}\,s^{-1}}$
and a line center at 1.92\,\AA. The latter value agrees within the
uncertainties with the wavelength of the  Fe~K$\alpha$
line (1.94\,\AA). Hence we conclude that cool-wind material 
must be irradiated by X-rays which are hard enough ($>7.1$\,keV) 
to excite this line by fluorescence. Our PoWR wind models show that 
in the outer wind the leading ionization stage  of iron is Fe\,{\sc v}. 
The presence of fluorescence is consistent with a two-component medium, 
where the cool wind  is permeated with the hot X-ray emitting plasma.   

\section{The cool wind model of WR\,6}

To determine the physical conditions in the ``cool'' component of the 
stellar wind, we employ the PoWR model atmosphere code \citep{wrh2004}. 
The PoWR  code solves the non-LTE radiative transfer simultaneously 
with the equations of statistical and radiative equilibrium. Complex model 
atoms with thousands of transitions are taken into account. The extensive 
inclusion of the iron group elements is important because of their blanketing 
effect on the atmospheric structure \citep{graf2002}. A particular stellar 
atmosphere model is defined by the effective temperature, surface gravity, 
luminosity, mass-loss rate, wind terminal velocity, and chemical composition.

For the supersonic part of the wind, we adopt the radial dependence
of velocity as $v(r)=v_\infty(1-1/r)$, where $r$ is in
units of the stellar radius $R_\ast$, and the terminal velocity
$v _\infty$ is a free parameter. In the subsonic region, the velocity
field is defined such that the hydrostatic density stratification is
approached. The PoWR models account for stellar-wind clumping in
the standard volume-filling factor `microclumping' approximation
\citep{hil1991,wrh1998}. Our best model of WR\,6 has a clumping factor $D=20$,
where $D=\langle \rho^2\rangle/\langle \rho \rangle^2$. The
synthetic spectra are calculated over the whole spectral range from
UV to IR, and then compared to the observed spectra and photometric fluxes.

With the final wind model being adopted, the stratifications of the
density,  opacity,  ionization, and radiative flux are defined.  
The mass absorption increases with distance from the
star, partly because the fully ionized helium recombines to \eli{He}{2} 
with its strong bound-free opacity. This increase of the mass absorption
coefficient partly compensates for the decrease of matter density, keeping
the wind opaque to large radii.

\begin{figure*}
\centering
\includegraphics[width=0.7\columnwidth]{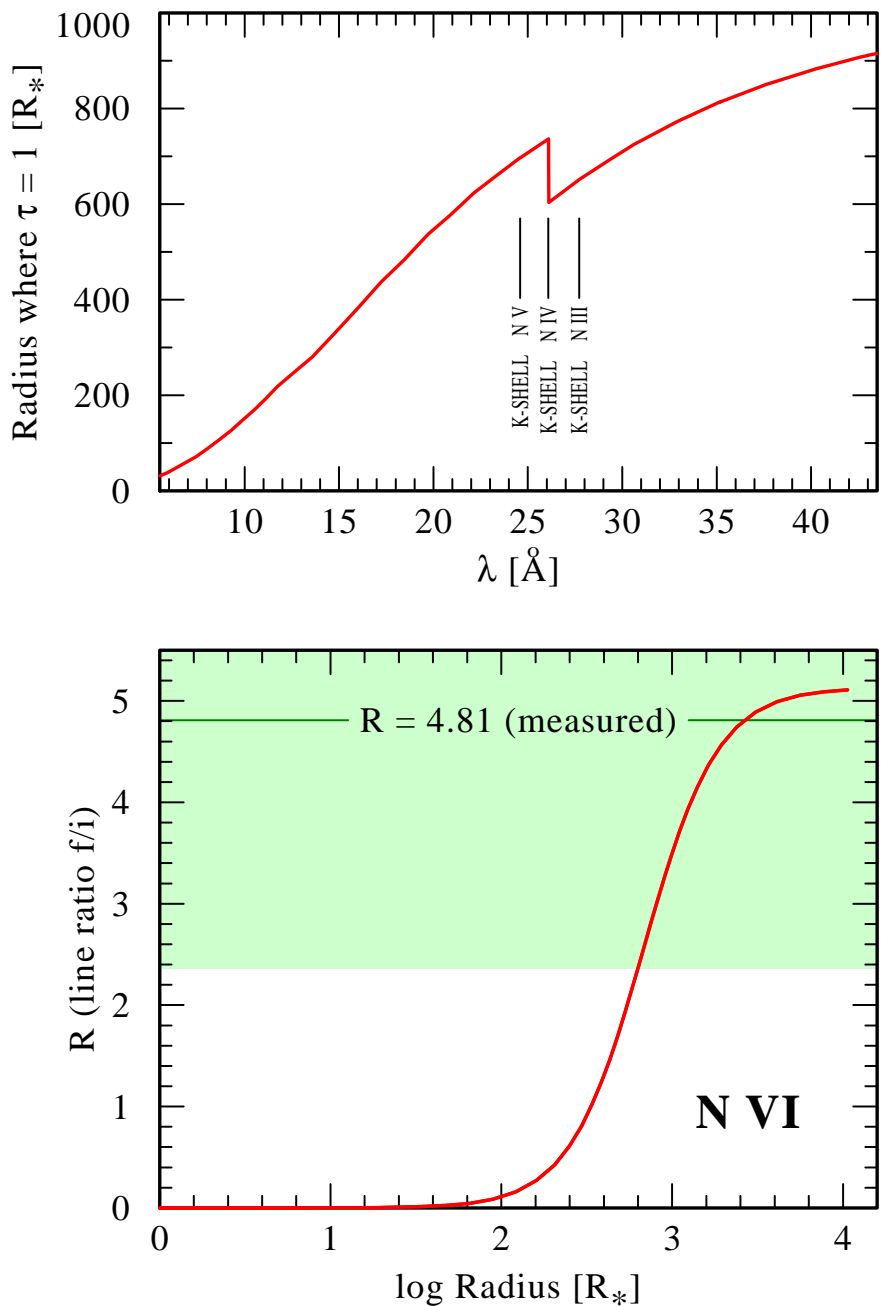}
\caption{{\it Upper panel}: radius where the continuum optical depth reaches 
unity, as predicted by the PoWR model for the ``cool'' wind  component 
of WR\,6, when inhomogeneities are only accounted for in the 
``microclumping'' approximation. {\it Lower panel}: 
Dependence of the line ratio $R=f/i$  for the N\,{\sc vi}
lines on the radial location  of the emitting plasma. 
 The measured value 
is shown as horizontal green line, with the green shaded area
representing the $3\sigma$ confidence band of the  measurement. }  
\label{fig:tau1} 
\end{figure*}

In Fig.\,\ref{fig:tau1} (upper panel) we show the radius in the 
wind where the continuum optical depth on the radial ray becomes unity, 
as predicted by PoWR model. Roughly speaking, only  X-rays emitted  
from outside that radius can escape from the wind and be seen by a distant 
observer, unless inhomogeneities create some porosity (``macroclumping''). 

\section{The analysis of line ratios in He-like ions}

\begin{figure*}
\centering
\includegraphics[width=0.7\columnwidth]{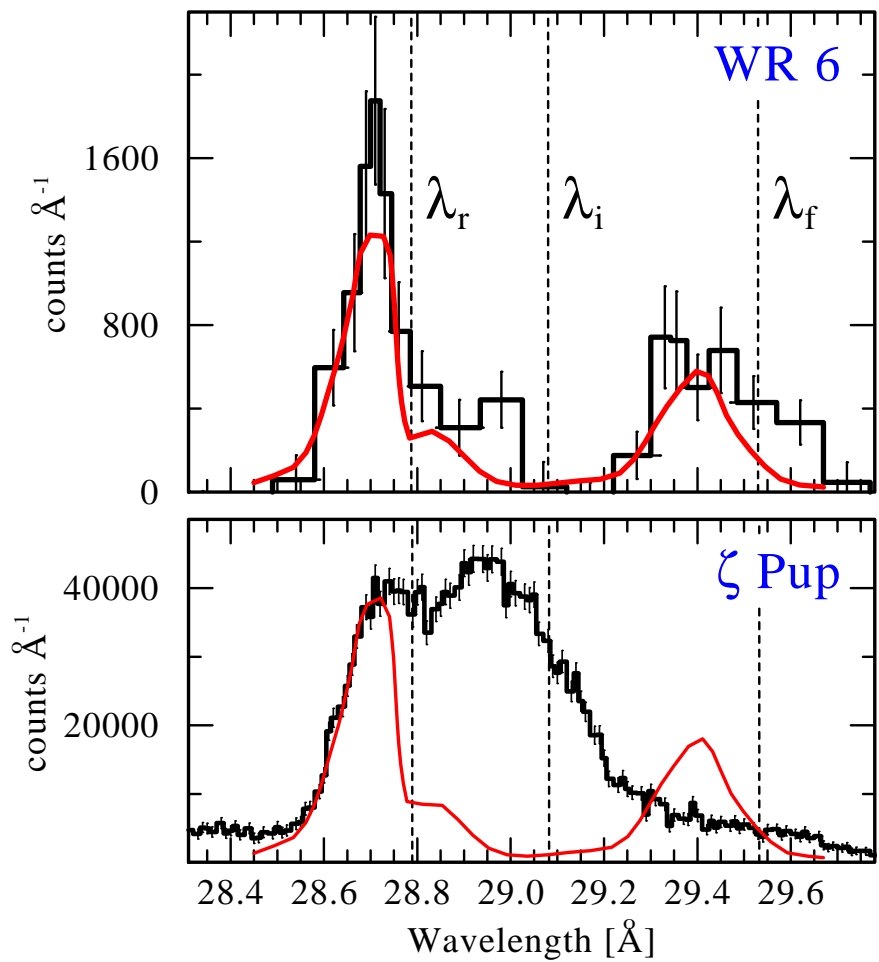}
\caption{{\em Upper panel:} RGS spectrum of WR\,6 in the range of   
the N\,{\sc vi} triplet of lines (black). The vertical error bars correspond to 
3$\sigma$. The vertical lines indicate the restframe wavelength of the resonance 
($\lambda_{\rm r}$), intercombination ($\lambda_{\rm i}$), and forbidden 
($\lambda_{\rm f}$) line, respectively. The solid red line shows a fit to these 
lines by  a formal model with three Gaussian profiles. 
{\em Lower panel:}  Same as upper, but for the O4I star $\zeta$\,Puppis. 
The red line shows the scaled  Gaussian fit from the upper panel.}
\label{fig:nvi} 
\end{figure*}

Helium-like ions  show a group of three
lines, consisting of a forbidden ($f$), an intercombination ($i$), and
a resonance (r) transition -- the so-called $fir$ triplet. 
The $fir$ triplets of \eli{Si}{13}, \eli{Mg}{11}, \eli{Ne}{9}, and \eli{N}{6} 
are present in the RGS spectrum of WR\,6. As
illustrated in Fig.\,\ref{fig:nvi}, the line ratios observed in WR\,6 are very 
different from those observed in typical single O-type stars.  
For N\,{\sc vi} the ratio $f/i$ exceeds unity in WR\,6, while O stars 
show $f/i<1$ \citep{leu2006,waldron2007}.

For each He-like ion,   the ratio of fluxes between
the forbidden and the intercombination component,  $R$, is sensitive to
the electron density and the ultraviolet flux \citep{blum1972}:
\begin{equation}
 R(r)=\frac{{\cal R}_0}{1+\phi(r)/\phi_{\rm c}+N_{\rm e}(r)/N_{\rm c}},
\label{eq:r}
\end{equation}
where $\phi$ is the photoexcitation rate from the term 2s\,$^3$S
to 2p$^3$P, and $N_{\rm e}$ is the electron density. The
quantities ${\cal R}_0, \phi_{\rm c}$, and $N_{\rm c}$ depend only
on atomic parameters and the electron temperature \citep{blum1972,porq2001}.

We use the PoWR model to calculate the values of $R(r)$ for
\eli{Mg}{11},  \eli{Ne}{9}, and \eli{N}{6} as function of the radial
location in the  wind of WR\,6. The density in the X-ray emitting shocks is not
known, therefore we  computed two sets of $R(r)$, one neglecting the
density term, and another one  assuming that the density in the shock
is the same as in the ambient ``cool'' wind but the plasma is fully ionized. 
According to the strong-shock condition, the hot gas density would be even 
a factor of four higher than that \citep{zel1966}. The photo-excitation  rates
$\phi(r)$ are computed at each radius from the   radiation intensity as
provided by the PoWR model,  which accounts not only simply for
geometric dilution, but  also for the diffuse radiative field.
The model shows that radiative depopulation of the metastable level 
dominates over collisional depopulation. Fig.\,\ref{fig:tau1} (low panel) 
shows for the N\,{\sc vi} triplet the predicted $R$ ratio  as function of 
the radial location of the emitting plasma. The measured $R=f/i$ ratio for 
the N\,{\sc vi}  is only compatible with the model if the X-ray emitting 
plasma is more than $\approx 630\,R_\ast$ away from the star. The 
best-fit value of the line ratio $R$ deduced from observations 
corresponds to the radial location of the emitting plasma at 
$\msim 2000\,R_\ast$. 
The  N\,{\sc vi} emission  line indicates a plasma temperature of  
$\approx 1.58$\,MK, which could be reached in a strong  shock with a 
velocity jump by $\approx 400$\,km\,s$^{-1}$.

\section{Modeling of X-ray emission line profiles}

\begin{figure*}
\centering
\includegraphics[width=0.7\columnwidth]{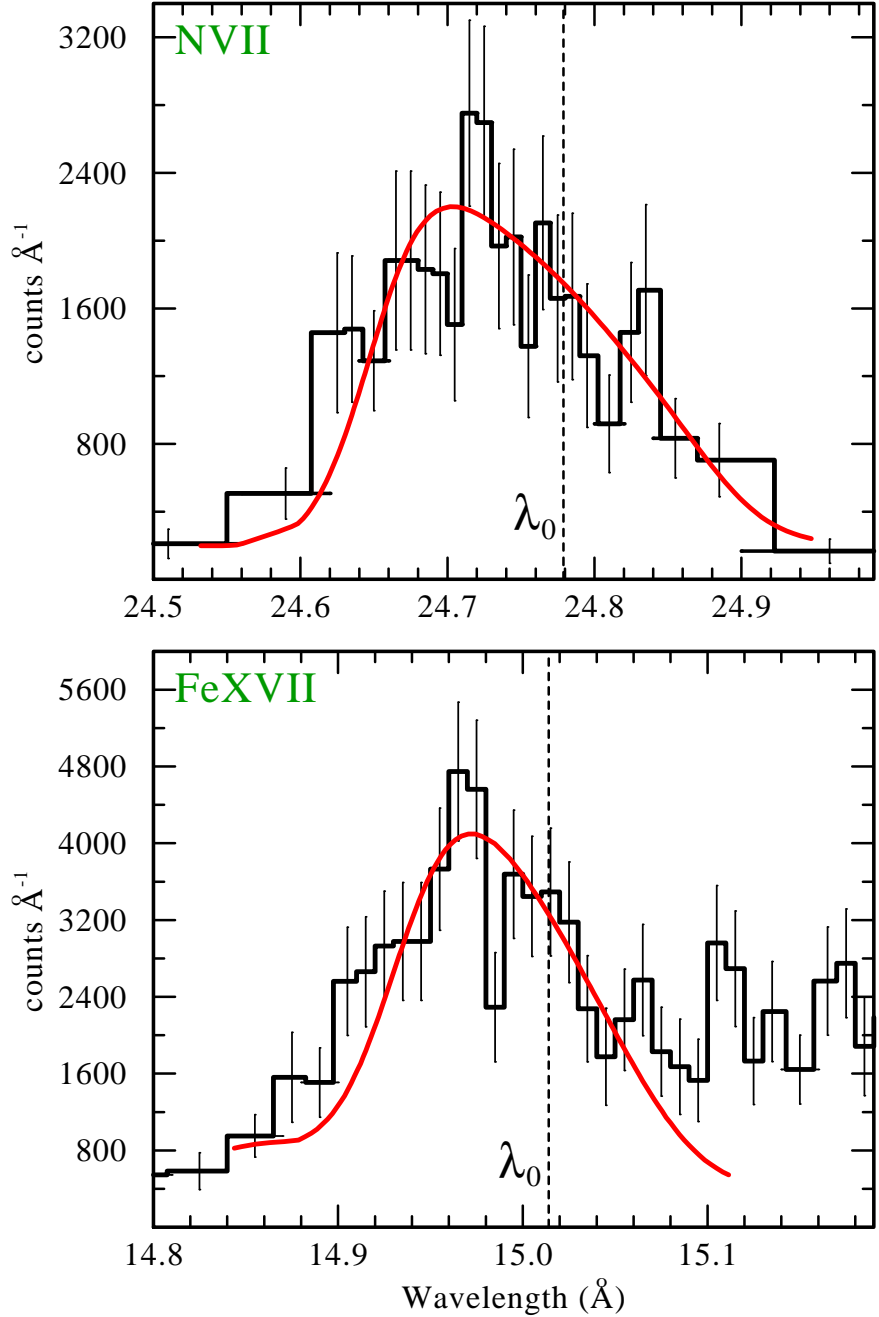}
\caption{{\em Upper panel:} Part of the spectrum of WR\,6 centered  on
the N\,{\sc vii} Ly$\alpha$ line. The observed spectrum is shown as a 
histogram (black). The
laboratory  wavelength ($\lambda_0$) of N\,{\sc
vii}\,Ly$\alpha\,\lambda$\,24.779\,\AA\  is indicated by a vertical
dashed line. The model line (smooth red) is a blend  of
N\,{\sc vii}\,Ly$\alpha$ $\lambda$\,24.779\,\AA\ and 
$\lambda$\,24.785\,\AA\ with a ratio  of line emissivities 1\,:\,0.5.
{\em Lower panel:} The same as upper  panel, but for Fe\,{\sc xvii}
$\lambda 15.014$\,\AA. The laboratory wavelength
($\lambda_0=15.014$\,\AA) is indicated by a  dashed  line. The
model line (smooth red)  is a blend of Fe\,{\sc
xvii}\,$\lambda$\,15.014\,\AA\ and $\lambda$\,14.892\,\AA\ with a ratio
of line emissivities 1\,:\,0.1. } 
\label{fig:nvii} 
\end{figure*}

The broadening of the X-ray lines from WR\,6 is consistent with the  known
terminal velocity of the wind of $\approx 1700$\,km\,s$^{-1}$.  The
lines centers are displaced from  their laboratory wavelengths  by
$\approx -0.06$\,\AA\  (see Fig.\,\ref{fig:nvii}).  Such
blue-shifts are predicted if the line emission is distributed within a
partly absorbing wind \citep[e.g.][]{macf1991,ign2001}. However, according to the
standard model, the wind of WR\,6 remains optically thick till large
distances from the star (550 -- 900\,$R_\ast$ for $\lambda > 20$\,\AA). 
X-rays can hardly emerge from interior to this location unless the wind 
is very porous.  

We employed our 2-D stochastic wind code \citep{osk2004} 
to  model the observed X-ray lines. Relaxing the
microclumping approximation, the clumps may have arbitrary optical
depth (``macroclumping'').  This wind fragmentation alters the
radiative transfer drastically, compared to a homogeneous wind with 
the same mass-loss rate \citep{feld2003}. The X-rays can escape from 
deeper in the wind than indicated in the upper panel in Fig.\,\ref{fig:tau1} 
for the smooth-wind model. 

We assume that parcels of  X-ray emitting gas are distributed
between clumps of continuum-absorbing, cool wind. The wind is fully 
fragmented throughout the whole relevant  radial range.  The  X-ray 
emission is placed only at radii larger than $r_{\rm em}$.  Since 
thermal X-ray emission arises from the decay of collisionally excited 
levels and  from free-free emission, is plausible to assume that 
the emissivity  scales as the square of the density,  
$\eta_\lambda \propto \rho^2$. With this scaling, most
X-rays are released close to $r_{\rm em}$. The intensity along any ray
is reduced by the factor $\exp({-\tau^{\rm c}_\lambda})$, where 
$\tau^{\rm c}_\lambda$ is the summation over the optical depths 
through all individual clumps which are encountered.

A grid of line  models was calculated for different values of 
the radius  $r_{\rm em}$ and $\left\langle N_{\rm c} \right\rangle$ -- 
the time-averaged number of fragments in a radial direction.
The radial dependence of the mass-absorption coefficient and the wind 
opacity  was accounted for in the models. Even with porosity, emission 
arising from too close to the star is totally trapped, while the adopted 
scaling of the emissivity with $\rho^2$ requires an unrealistic amount of 
energy being converted into X-rays. This excludes small values of 
$r_{\rm em}$. In the other extreme, when X-ray emission starts only at 
radii where the wind is already nearly transparent, the observed blue-shift 
and asymmetry of the line profiles cannot be reproduced. The model line 
profiles shown in Fig.\,\ref{fig:nvii} were calculated with 
$r_{\rm em} = 30\,R_\ast$ and $\left\langle N_{\rm c} \right\rangle = 50$. 

\section{On The Origin of X-ray Emission in WR stars }

Observed temperatures up to 50\,MK require shock velocity jumps comparable
to the wind speed. From UV spectra of WR\,6 there is  direct evidence  
that a fraction of the wind flows with much higher velocity  (up to 
3,100\,km\,s$^{-1}$) than the general wind speed (1,700\,km\,s$^{-1}$)
\citep{prinja1990}.   We suggest that the hottest gas is produced when
very fast flows impacts on slowly moving clumps within
the wind acceleration zone.  Such extreme shocks cannot be maintained to
large radii, but the hardest X-rays can escape from the inner wind owing to
smaller absorption cross-sections at higher energies.

Exceptionally large, pancake-shaped clumps which form close to the
photosphere at low velocities would resist the radiative acceleration. 
Being an obstacle to the generally much faster wind, they   
create strong reverse shocks that may add more mass to the clump if
the gas cools quickly, or generate bow shocks if it does not
\citep{cas2008}.  Self-consistent kinematic models of Gayley (in prep.) 
show that 500\,km\,s$^{-1}$ shocks may be present at $\sim 50\,R_\ast$
in the wind, suitable for X-ray escape if the wind is sufficiently
porous. 

\section{On the binary hypothesis for WR\,6}

The X-ray emission of WR\,6 varies considerably \citep{pol1987,wil1996}. 
Our new observations confirm and extend these results.  The \xmm\ light 
curve shows different variability time scales, including a variation 
of 20\% over a month's span. We also observe some spectral changes. 
The X-ray variability does not follow the 3.766\,d period which is seen 
in the optical and UV \citep{wil1989}. The X-ray variability of WR\,6  
further highlights the differences between single WR and O stars. The 
latter are remarkably constant X-ray sources  (Naz\'e \etal\ 2012, submitted).

The optical variability of WR\,6  has long been known \citep{wilson1948}, 
and it appears to be incompatible with stellar pulsations or the Be star 
phenomenology  \citep{rob1992}. Most plausibly, the variability is due to 
changes of wind structures \citep{stlou1993,flores2011}. This explanation is 
in line with our new X-ray observations.

\citet{fir1980} established that if the 3.766\,day period were  associated with 
binarity, the mass of the secondary star could not be  larger than $\approx 1.5\,M_\odot$.  
A star of such mass and age (assuming coeval formation of both binary components)  
would be either  a degenerate neutron star (NS) or a pre main-sequence 
young stellar  object (T\,Tauri type star). 

The NS scenario  was criticized because of the relatively low X-ray 
luminosity of WR\,6 and the lack of evidence in its UV spectrum \citep{stev1988}.
The WR+T\,Tau scenario  is also extremely unlikely, as the high-resolution 
X-ray spectrum of WR\,6 is  incompatible with the X-ray properties of young 
stellar objects \citep{kas2002}. 

Variability is a common property of WR stars. Among rigorously
monitored  WN stars, 40\%\ show optical variability similar to WR\,6
\citep{chen2011}. The assumption that 40\%\ of all WN stars have a NS or 
a T\,Tauri type companion is not realistic. 

Our high-resolution X-ray spectra show that the X-ray emitting plasma
moves at about the same velocity as the cool wind.  The X-ray line 
blue-shifts do apparently  not change with time. The X-ray emission 
line spectrum is compatible with the WN star abundances. All
these facts add further evidence against WR\,6 being a binary system.

\section{Conclusions}

Assuming that WR\,6 is typical, our results show that in the winds of WN 
stars strong shocks are active at large radii from stellar photosphere. 
A possible explanation involves slow, dense clumps that accumulate from 
the wind close above the photosphere without being  effectively 
accelerated. The seeding of such  slow clumps might be associated with 
(sub)surface convective zones  \citep{can2011}. {\changed The effects of 
such wind anisotropies on the distribution of circumstellar matter could 
influence the behavior of SN\,Ibc.} 

In any case, from high-resolution X-ray spectral data 
of WR\,6 we conclude that  unknown mechanisms must operate in WR winds.  
Identifying these mechanisms poses a challenging problem for the theory 
of stellar winds.

\acknowledgments
Based on observations made with \xmm, an ESA 
science mission with instruments and contributions directly funded by 
ESA member states and the USA (NASA). We thank R.K.\ Smith and A.\,Foster 
for help in computing emissivities for a hydrogen-depleted 
plasma. DPH was supported by NASA through the Smithsonian Astrophysical
Observatory contract SV3-73016 for the Chandra X-Ray Center and 
Science Instruments. Funding for this research has been provided by DLR grant 
50\,OR\,1101 (LMO).




%

\newpage

\newpage
%

\newpage
%

\newpage

\end{document}